\documentclass[12pt]{article}
\usepackage{amsmath,amssymb}

\newcommand{\bea}{\begin{eqnarray}}
\newcommand{\eea}{\end{eqnarray}}
\newcommand{\be}{\begin{equation}}
\newcommand{\ee}{\end{equation}}
\newcommand{\vs}[1]{\vspace{#1 mm}}

\newcommand{\dsl}{\pa \kern-0.5em /}

\newcommand{\pa}{\partial}

\newcommand{\nn}{\nonumber\\}

\begin{document}
\topmargin 0mm
\oddsidemargin 0mm

\begin{flushright}

USTC-ICTS-19-19\\

\end{flushright}

\vspace{2mm}

\begin{center}

{\Large \bf A note on the open string pair production of the D3/D1 system}

\vs{10}

{\large J. X. Lu}

\vspace{4mm}

{\em
Interdisciplinary Center for Theoretical Study\\
 University of Science and Technology of China, Hefei, Anhui
 230026, China\\
 
}

\end{center}

\vs{10}

\begin{abstract}
    We here report a possibility of detecting the open string pair production from the system of a D3 brane, taken as our own (1 + 3)-dimensional world and carrying its worldvolume collinear electric and magnetic fluxes, and a nearby D-string, placed parallel at a separation  in the directions transverse to both branes. 
    \\
\end{abstract}

\newpage

In a series of recent publications by the present author and his collaborators \cite{Lu:2017tnm, Lu:2018suj, Lu:2018nsc, Jia:2018mlr}, we uncover the existence of  open string pair production, in the spirit of Schwinger pair production in QED \cite{Schwinger:1951nm},  for a system of two Dp branes, placed parallel at a separation, with each carrying electric and magnetic fluxes  in Type II superstring theories.  The open string pair production for un-oriented  bosonic string or un-oriented Type I superstring was discussed a while ago by Bachas and Porrati in \cite{Bachas:1992bh, Porrati:1993qd}.

Our studies indicate that a significantly enhanced open string pair production can be achieved when the applied magnetic field strength does not share any common field strength index with the electric one and as such we need to have $p \ge 3$ in general \cite{ Lu:2017tnm, Lu:2018suj}. Among these,  the $p = 3$ system actually gives rise to the largest open string pair production rate if the applied electric and magnetic fluxes remain the same for all $p \ge 3$.  Adding more magnetic fluxes in a similar manner  reduces rather than increases the pair production rate \cite{Jia:2018mlr}.  So the largest possible rate is for the $p = 3$ system when the applied electric field and the magnetic one are collinear. 

Given this, a natural question arises: can this effect be utilized in a way to test the underlying string theory if we take one of the D3 branes as our own (1 + 3)-dimensional world?   This may open a new possibility of testing the underlying string theory without the need to compactify the higher dimensional string/M theory to give the 4-dimensional particle physics standard model, for example. A byproduct of this, from the view of a worldvolume observer, is to find the existence of extra dimensions. 

However, for the simplest system of two D3 branes mentioned above, our recent study given in \cite{Lu:2018nsc} finds this unlikely.  One of reasons is that the current  laboratory controllable electric and magnetic fields are too small for this purpose. The other, actually related, is that the underlying system is still 1/2 BPS in the absence of worldvolume fluxes and this gives rise to the mass scale of the lowest massive modes of the open string no less than a few TeV if not larger.   So finding a way to lower the mass scale may provide a solution to the question raised, for example, by  considering a non-supersymmetric system rather than a supersymmetric one from the outset.  

Does there exist indeed such a suitable system in the underlying string theory to make the detection of so produced open string pair production a potential possibility?  We try to address this in the present paper.

It is well-known that a magnetic flux in a D-brane is equivalent to co-dimensional 2 D-branes delocalized along the flux field strength directions inside the original D-brane \cite{Breckenridge:1996tt, Costa:1996zd, Di Vecchia:1997pr}.  A while ago, the present author and his collaborator pointed out \cite{Lu:2009pe} that a Dp$'$ brane, with $p - p' = 2$,  behaves effectively like a pure string-scale magnetic field  to a Dp brane, in the discussion of open string pair production.  We just made the observation there but didn't realize its significance or usefulness then.  Moreover,  the system of a Dp and a Dp$'$, with $p - p' = 2$ and placed parallel at a separation, does not preserve any supersymmetry, even in the absence of  worldvolume fluxes.  So this kind of system appears just to serve our need mentioned earlier.  In a very recent paper \cite{Jia:2019hbr},  we demonstrated that the largest open string pair production rate for practically small electric field on the branes occurs for the system of a D3 and a D1, placed parallel at a separation in the directions transverse to both branes, with the D3 carrying both an electric flux and a collinear magnetic one while the D1 carrying a collinear different electric flux. 

We will focus on this particular system in this paper and use it as an explicit example to demonstrate the potential possibility of detecting the open string pair production so produced.

The closed string tree-level cylinder amplitude between the D3 and the D1 can be read from the last equality of the general one given in  (131) in \cite{Jia:2019hbr} for $p = 3, p' = 1$ as
\bea\label{c-ampli}
&&\Gamma_{3, 1} = \frac{2 V_{2} \sqrt{(1 - \hat f'^{2})(1 - \hat f^{2})(1 + \hat g^{2})} (\cosh \pi \bar\nu_{0} - \cos\pi \nu_{1})^{2}}{8 \pi^{2} \alpha'} \int_{0}^{\infty} \frac{d t}{t^{3}} ~e^{- \frac{y^{2}}{2 \pi \alpha' t}} \nn
&&\times \prod_{n = 1}^{\infty} \frac{[1 - 2 |z|^{2n} e^{-\pi \bar\nu_{0}} \cos\pi\nu_{1} + e^{- 2\pi \bar\nu_{0}} |z|^{4n}]^{2} [1 - 2 |z|^{2n} e^{\pi \bar\nu_{0}} \cos\pi\nu_{1} + e^{ 2\pi \bar\nu_{0}} |z|^{4n}]^{2} }{(1 - |z|^{2n})^{4} [1 - 2 |z|^{2n} \cosh2\pi \bar\nu_{0} 
+ |z|^{4n}] [1 - 2 |z|^{2n} \cos2\pi\nu_{1} 
+ |z|^{4n}]},
\eea
where $y$ is the brane separation along the directions transverse to both D3 and D1, $|z| = e^{- \pi t} < 1$, $V_{2}$ the volume of D1 worldvolume, $\alpha'$ the Regge slope parameter, and the so-called electric parameter $\bar\nu_{0}$ and the magnetic parameter $\nu_{1}$ are determined by the electric fluxes ($\hat f, \hat f'$) and the magnetic one ($\hat g$), respectively, via
\be\label{nu-parameter}
\tanh \pi \bar\nu_{0} = \frac{|\hat f - \hat f'|}{1 - \hat f \hat f'}, \qquad \tan \pi \nu_{1} = \frac{1}{|\hat g|},
\ee
with $\bar\nu_{0} \in [0, \infty), \nu_{1} \in (0, 1/2]$.  In the above, 
the D3 worldvolume flux $\hat F_{3}$ and the D1 electric flux $\hat F'_{1}$ are given, respectively, as
\be\label{flux}
\hat F_{3} = \left(\begin{array}{cccc}
0& \hat f&0&0\\
-\hat f&0&0&0\\
0&0&0&\hat g\\
0&0&- \hat g&0\end{array}\right), \qquad \hat F'_{1} = \left(\begin{array}{cc}
0 & \hat f'\\
- \hat f'& 0\end{array}\right)
\ee
Note that the dimensionless flux (denoted with a hat above) here is defined via $\hat F = 2\pi \alpha' F$ with $F$ the usual dimensionful field strength (without a hat above).  So we have $|\hat g| \in [0, \infty)$ and $|\hat f|, |\hat f'| \in [0, 1)$ with unity here as the  critical value of the respective electric flux. Note also that in the absence of worldvolume fluxes (therefore $\bar \nu_{0} = 0, \nu_{1} = 1/2$ from (\ref{nu-parameter})), the above amplitude is strictly positive, giving an attractive interaction between the D3 and the D1 according to our conventions. This indicates that the underlying system does not preserve any supersymmetry, consistent with the known fact. For large $y$, the dominant contribution to the amplitude comes from the large $t$ integration, due to the exponentially suppressing factor ${\rm Exp} [- y^{2} /(2\pi\alpha' t)]$ in the integrand, and the amplitude is therefore positive since every factor in the integrand is positive, giving also an attractive interaction.  For small $y$, the small $t$ integration becomes important and the factor $[1 - 2 |z|^{2n} \cosh 2\pi \bar \nu_{0} + |z|^{4n}] \approx 2 (1 - \cosh 2\pi\bar\nu_{0}) $  in the denominator of the infinite product in the integrand of (\ref{c-ampli}) can be negative for small $t$.  Once this happens, the sign of the amplitude is ambiguous since there are an infinite number of such factors in the product.  This ambiguity indicates a potential new physics to occur and the best way to decipher this is to pass the closed string cylinder amplitude to the corresponding open string one-loop annulus one via a Jacobi transformation\footnote{Certain relations for the Dedekind $\eta$-function and the $\theta_{1}$-function have also been used, see \cite{Jia:2019hbr} for detail, for example.} by sending $t \to 1/t$. The resulting annulus amplitude is
\bea\label{annulus-amplit}
\Gamma_{3, 1} &=& \frac{2 V_{2} |\hat f - \hat f'| }{8 \pi^{2} \alpha'} \int_{0}^{\infty} \frac{d t}{t} ~e^{- \frac{y^{2} t}{2 \pi \alpha' }} \frac{[\cosh \pi\nu_{1} t - \cos\pi\bar\nu_{0} t]^{2}}{\sin\pi\bar\nu_{0} t \sinh\pi\nu_{1} t}\nn
&\,& \times \prod_{n = 1}^{\infty} \frac{|1 - 2 |z|^{2n} e^{- i \pi \bar\nu_{0} t} \cosh \pi \nu_{1} t  + |z|^{4n} e^{- 2 i \pi \bar\nu_{0} t}|^{4}}{(1 - |z|^{2n})^{4} [1 - 2 |z|^{2n} \cosh 2\pi \nu_{1} t + |z|^{4n}][1 - 2 |z|^{2n} \cos2\pi\bar\nu_{0} t + |z|^{4n}]},
\eea
where $|z| = e^{- \pi t}$ continues to hold.  Except for the factor $\sin\pi\bar\nu_{0} t$, all other factors in the integrand are positive since $1 - 2 |z|^{2n} \cos\pi\nu_{1} t + |z|^{4n} > 1 - 2 |z|^{2n} + |z|^{4n} = (1 - |z|^{2n})^{2} > 0$ and 
$1 - 2 |z|^{2n} \cosh2\pi \nu_{1} t + |z|^{4n} = (1 - e^{2\pi \nu_{1} t} |z|^{2n}) (1 - e^{- 2\pi \nu_{1} t} |z|^{2n}) > 0$ since $\nu_{1} \le 1/2$ and $n \ge 1$.  The factor $\sin\pi\bar\nu_{0} t$ actually gives rise to an infinite number of simple poles of the integrand at $t_{k} = k/\bar\nu_{0}$ with $k = 1, 2, \cdots$ along the positive t-axis, reflecting the existence of an imaginary part of the amplitude.  This imaginary part indicates the decay of the underlying system via the so-called open string pair production.  The decay rate can be computed as the sum of the residues of the integrand in (\ref{annulus-amplit}) at these poles times $\pi$ per unit worldvolume following  \cite{Bachas:1992bh} as
\be\label{decay-rate}
{\cal W} = \frac{4 |\hat f - \hat f'|}{8 \pi^{2} \alpha'}\sum_{k = 1}^{\infty} \frac{(- )^{k - 1}}{k} e^{- \frac{k y^{2}}{2\pi \alpha' \bar\nu_{0}}} \frac{\left[\cosh \frac{\pi k \nu_{1} }{\bar\nu_{0}} - (-)^{k}\right]^{2}}{\sinh \frac{\pi k \nu_{1} }{\bar\nu_{0}}} Z_{k} (\bar\nu_{0}, \nu_{1}),
 \ee
where \be\label{Z}
Z_{k} (\bar\nu_{0}, \nu_{1}) = \prod_{n =1}^{\infty} \frac{\left[1 - 2 (-)^{k} |z_{k}|^{2n} \cosh \frac{\pi k \nu_{1} }{\bar\nu_{0}} + |z_{k}|^{4n} \right]^{4}}{(1 - |z_{k}|^{2n})^{6} \left[1 - 2 |z_{k}|^{2n} \cosh \frac{2\pi k \nu_{1} }{\bar\nu_{0}} + |z_{k}|^{4n}\right]},
\ee
with $|z_{k}| = e^{- \pi k/\bar\nu_{0}}$.
Following \cite{nikishov},  the rate for the open string pair production corresponds just to the leading $k = 1$ term of the above decay rate and it is 
\be\label{pprate}
{\cal W}^{(1)} = \frac{4 |\hat f - \hat f'|}{8 \pi^{2} \alpha'} e^{- \frac{y^{2}}{2\pi \alpha' \bar\nu_{0}}} \frac{\left[\cosh \frac{\pi  \nu_{1} }{\bar\nu_{0}} + 1\right]^{2}}{\sinh \frac{\pi  \nu_{1} }{\bar\nu_{0}}} Z_{1} (\bar\nu_{0}, \nu_{1}),
\ee
where 
\be\label{Z1}
Z_{1} (\bar\nu_{0}, \nu_{1}) = \prod_{n =1}^{\infty} \frac{\left[1 + 2  |z_{1}|^{2n} \cosh \frac{\pi  \nu_{1} }{\bar\nu_{0}} + |z_{1}|^{4n} \right]^{4}}{(1 - |z_{1}|^{2n})^{6} \left[1 - 2 |z_{1}|^{2n} \cosh \frac{2\pi  \nu_{1} }{\bar\nu_{0}} + |z_{1}|^{4n}\right]}.
\ee
With the above computed open string pair production rate, we can now discuss the possibility of detecting the pair production. Suppose that the D3 is the (1 + 3)-dimensional world we are living in and there exists a nearby D1 in the directions transverse to our D3.
For practical purpose, we can only control the sizes of the collinear electric and magnetic fields on our D3, both of which are in general  very small compared with the string scale. In other words,  $|\hat f| \ll 1$ and $|\hat g| \ll 1$ and for simplicity we set 
$\hat f' = 0$ from now on.  With these, we have from (\ref{nu-parameter}), $\bar\nu_{0} \approx |\hat f /\pi| \ll 1$ and $\nu_{1} \lesssim 1/2$.  So we have $Z_{1} (\bar\nu_{0}, \nu_{1}) \approx 1$, noting $|z_{1}| = e^{- \pi/\bar\nu_{0}}\to 0$, and $\nu_{1}/\bar\nu_{0} \gg 1$. The pair production rate (\ref{pprate}) then becomes 
\be\label{pprate-new}
{\cal W}^{(1)} \approx \frac{e E}{2 \pi }\, e^{- \frac{\pi m^{2}_{\rm eff}}{e E}} ,
\ee
where we have used $|\hat f| = 2\pi \alpha' e E$ with $e$ the charge carried by positron and $E$ the laboratory electric field, and we have defined an effective mass  $m_{\rm eff}$ with $m^{2}_{\rm eff} = m^{2} - \nu_{1}/(2\alpha')$ where we have  introduced the string scale $m = T_{f} y = y/(2\pi\alpha')$ with $T_{f}$ the fundamental string tension. The magnetic flux $\hat g = 2\pi \alpha' e B \ll 1$ with $B$ the usual laboratory magnetic field.  Note that the exponential factor ${\rm Exp}[\pi \nu_{1}/(2\alpha' e E)]$ appearing in the above rate, due to the so-called tachyonic shift which will be discussed later on, can be very large for the present system, therefore enhances this rate significantly.

In terms of the effective mass $m_{\rm eff}$, the above rate resembles\footnote{There is a difference of factor $e E/(2\pi)$  between the two rates due to the presence of D1 for the present case.}, to certain extent,  the Schwinger pair production one for electron and positron, in the absence of magnetic field,  for which \cite{Schwinger:1951nm}
 \be\label{schwingerrate}
 {\cal W}_{\rm QED} = \frac{(e E)^{2}}{(2\pi)^{2}}\,  e^{- \frac{\pi m^{2}}{e E}}.
 \ee
Before we address the usefulness of the present rate (\ref{pprate-new}), let us make it clear which stringy modes actually contribute to the rate.  To make the discussion easy to follow, we adopt the trick used in \cite{Jia:2019hbr} by taking D1 effectively as a D3 carrying an infinitely  large magnetic flux along 2 and 3 directions in the following way
\be\label{flux}
\hat F'_{3} = \left(\begin{array}{cccc}
0& 0&0&0\\
0&0&0&0\\
0&0&0&\hat g'\\
0&0&- \hat g'&0\end{array}\right),
\ee
with $\hat g' \to \infty$.  For the moment, let us take $\hat g'$ to be large.  We have now 
\be\label{general-nu1}
\tan\pi\nu_{1} = \left|\frac{\hat g - \hat g'}{1 - \hat g \hat g'}\right|,
\ee
where the second equation in (\ref{nu-parameter}) is the limit of the above when we take $\hat g' \to \infty$.  We can follow the discussion given in \cite{Lu:2018nsc}.  In the absence of worldvolume fluxes,  the mass spectrum of the open superstring connecting the two D3 is given as
 \be\label{mass-level}
\alpha' M^{2} = -\alpha' p^{2} =  \left\{\begin{array}{cc}
\frac{y^{2}}{4\pi^{2} \alpha'} + N_{\rm R} &  (\rm R-sector),\\
\frac{y^{2}}{4\pi^{2} \alpha'}  + N_{\rm NS}  - \frac{1}{2} & (\rm NS-sector),
\end{array}\right.
\ee
where $p = (k, 0)$ with $k$ the momentum along the brane worldvolume directions, $N_{\rm R}$ and $N_{\rm NS}$ are the standard number operators in the R-sector and NS-sector, respectively, as
\bea\label{NP}
N_{\rm R} &=& \sum_{n = 1}^{\infty} (\alpha_{- n} \cdot \alpha_{n} + n d_{-n} \cdot d_{n}),\nn
N_{\rm NS} &=& \sum_{n = 1}^{\infty} \alpha_{- n} \cdot \alpha_{n} + \sum_{r = 1/2}^{\infty} r d_{- r} \cdot d_{r}.
\eea
When $y = 0$, the supersymmetric open string has 16 massless modes, which are eight bosons ($8_{\rm B}$) and eight fermions ($8_{\rm F}$), in addition to an infinite tower of massive modes.  In terms of the D3 brane worldvolume description, these 16 massless modes correspond to one of U(2) generators of the 4-dimensional N = 4 super Yang-Mills, which is broken when $y \neq 0$.  The massless $8_{\rm B}$ consists of a massless vector and 6 massless scalars in four dimensions when $y = 0$ and gives a massive vector, the W-boson, plus 5 massive scalars when one of 6 massless scalars takes a non-vanishing vacuum expectation, giving $y \neq 0$. The $8_{\rm F}$ gives four 4-dimensional Majorana spinors.  We have now massive $8_{\rm B} + 8_{\rm F}$, all charged under the respective unbroken U(1) and all with the same mass $T_{f} y = y/(2\pi \alpha')$ as given in (\ref{mass-level}).  The latter is consistent with the fact that the underlying supersymmetry remains the same before and after the gauge symmetry breaking.
 
Let us now consider to turn on only a small magnetic fluxe $\hat g \ll 1$ on our own D3 and a large $\hat g'$ on the other D3 which is taken effectively as the D1 when $\hat g' \to \infty$ mentioned earlier.  The energy in the R-sector is strictly positive and the R-sector vacuum, consisting of massive $8_{\rm F}$ or 4 massive four-dimensional Majorana spinors when $y \neq 0$, has no the so-called tachyonic shift, just like the QED case \cite{Ferrara:1993sq,Bolognesi:2012gr}.  
The NS-sector has a different story, however. The energy spectrum in the NS-sector, following \cite{Ferrara:1993sq,Bolognesi:2012gr}, is now
\be\label{OSM}
\alpha' E^{2}_{\rm NS} =  (2 N + 1) \frac{\nu_{1}}{2} - \nu_{1} \, S + \alpha' M^{2}_{\rm NS},
\ee
where the magnetic parameter $\nu_{1}$ is defined in (\ref{general-nu1}),  the Landau level  $N = b^{+}_{0} b_{0} $, the spin operator in the 23-direction is 
\be\label{spin}
S = \sum_{n = 1}^{\infty} (a^{+}_{n} a_{n} - b^{+}_{n} b_{n}) + \sum_{r = 1/2}^{\infty} (d^{+}_{r} d_{r} - \tilde d^{+}_{r} \tilde d_{r}),
\ee
and the mass $M_{\rm NS}$ in the NS-sector is
\be\label{new-mass}
\alpha' M^{2}_{\rm NS}  = \frac{y^{2}}{4 \pi^{2} \alpha'} + N_{\rm NS} - \frac{1}{2},
\ee
but now with  
\be
N_{\rm NS} = \sum_{n = 1}^{\infty}  n (a^{+}_{n} a_{n} + b^{+}_{n} b_{n}) + \sum_{r = 1/2}^{\infty} r (d^{+}_{r} d_{r} + \tilde d^{+}_{r} \tilde d_{r}) 
+ N^{\perp}_{\rm NS},
\ee
where $N^{\perp}_{\rm NS}$ denotes the contribution from directions other than the 2 and 3 ones.  The only states which have the potential to be tachyonic \cite{Ferrara:1993sq}, in particular for $y = 0$,  belong to the first Regge trajectory, given by
\be\label{fRt}
(a^{+}_{1})^{\tilde n} d^{+}_{1/2} |0 \rangle_{\rm NS},
\ee
where  $|0 \rangle_{\rm NS}$ is the NS-sector vacuum,  projected out in the superstring case. For these states, we have 
\be\label{EL}
\alpha' E^{2}_{\rm NS} = - \frac{\nu_{1}}{2} + (1 - \nu_{1}) (S - 1) +  \frac{y^{2}}{4 \pi^{2} \alpha'},
\ee
where $S = \tilde n + 1 \ge 1$.  In the absence of magnetic fluxes, the above gives the mass in the NS-sector as 
\be\label{ML}
\alpha' M^{2}_{\rm NS} = \frac{y^{2}}{4 \pi^{2} \alpha'} + S - 1,
\ee
which  becomes massless  for the spin $S = 1$ state $d^{+}_{1/2} |0\rangle_{\rm NS}$ when $y = 0$  and is massive for $y \neq 0$.    A finite brane separation $y$ for the spin $S = 1$ state $d^{+}_{1/2} |0\rangle_{\rm NS}$ corresponds to the gauge symmetry breaking $U(2)  \to U(1) \times U(1)$, with the W-boson mass $M_{W} = y/(2\pi \alpha')$,   as mentioned earlier.  

From (\ref{EL}) and (\ref{ML}), the lowest mass is for the spin $S = 1$ state and the state $d^{+}_{1/2} |0\rangle_{\rm NS}$ has its energy 
\be\label{energy}
\alpha' E^{2}_{\rm NS} = - \frac{\nu_{1}}{2} +  \frac{y^{2}}{4 \pi^{2} \alpha'},
\ee
and a so-called tachyonic shift $ \nu_{1}/2$ in the presence of magnetic flux\footnote{Adding an additional magnetic flux, say, along 45-direction with the corresponding magnetic parameter $\nu_{2} > 0$, without loss of generality assuming $\nu_{2} \le \nu_{1}$, will actually reduce rather than increase the shift to $(\nu_{1} - \nu_{2})/2$. So this will diminish rather than increase the rate as mentioned earlier.}.  This energy is precisely the effective mass defined for the rate (\ref{pprate-new}) given earlier. This same energy as well as the tachyonic shift can also be seen from the integrand of the open string annulus 
amplitude (\ref{annulus-amplit}) for the $t \to \infty$ limit, which gives a blowing-up factor $e^{- 2 \pi t [ - \frac{\nu_{1}}{2} + \frac{y^{2}}{4\pi^{2} \alpha'}] }$ for $y < \pi \sqrt{2 \alpha' \nu_{1}}$.  It is clear that this is due to the appearance of a tachyon mode in the presence of the magnetic flux.  Once this happens, we will have a phase transition via the so-called tachyon condensation \cite{Ferrara:1993sq}. From the viewpoint of the D3 worldvolume, this instability is the Nielsen-Olesen one for the non-abelian gauge theory of the 4-dimensional N = 4  U(2) super Yang-Mills\cite{Nielsen:1978rm} in the so-called weak field limit.

Unlike our previous discussion for the system of two D3 branes given in \cite{Lu:2018nsc},  we have here an intrinsic $\nu_{1} = 1/2$ even in the absence of the magnetic flux $\hat g$ on our own D3 brane for the present D3/D1 one for which the D1 is effectively taken as the $\hat g' \to \infty$ limit of the other D3 as mentioned earlier.   Precisely because of this finite non-vanishing value of $\nu_{1}$, we have a charged massive vector mode which is effectively lightest among the other charged massive ones originally from the $8_{\rm F} + 8_{\rm B}$.  As will be discussed later on, this finite value  provides us a potential possibility of detecting the open pair production if we add an electric flux $\hat f$ on our own D3 along the D1 spatial direction in practice. 
Note also that the present D3/D1 system preserves no supersymmetry even in the absence of added fluxes on the branes as mentioned earlier.

For the purpose of detecting the pair production,  we consider to add the collinear electric and magnetic fields on our own D3 as given in (\ref{flux}).  In practice, the present and near-future laboratory electric and magnetic fields are very small in comparison with the string scale which is at least on the order of a few TeV (for example, see  \cite{Berenstein:2014wva} as discussed in \cite{Lu:2018nsc}).  So we have $\bar \nu_{0} \ll 1$ and $\nu_{1} \lesssim 1/2$ from (\ref{nu-parameter}) in general.  As discussed earlier, $\bar \nu_{0} \ll 1$ implies $Z_{1} (\bar \nu_{0}, \nu_{1}) \approx 1$ and so we need to consider only the sixteen charged massive modes\footnote{Actually the other 16 modes carrying the opposite charge, due to the anti open string, are also included in the consideration} of $8_{\rm F} + 8_{\rm B}$ in the pair production rate (\ref{pprate}).  Further since $\nu_{1} /\bar\nu_{0} \gg 1$, the pair production rate is now given by (\ref{pprate-new}) which is due to the effectively lightest charged/anti charged massive vector modes.

The above discussion provides also a guidance for us to look for a possible test of the open string pair production rate (\ref{pprate-new})  in an earthbound laboratory.  In order to detect the pair production, for example, as an electric current produced by the ends of the produced open strings  as  charged particles on our own D3 brane, and not to violate the underlying unitary, we need  as usual  from (\ref{pprate-new}) to have 
\be\label{detectioncd}
e E \sim m^{2}_{\rm eff} = \frac{y^{2}}{(2\pi \alpha')^{2}} - \frac{\nu_{1}}{2 \alpha'} \ge 0,
\ee
where $\nu_{1} \lesssim 1/2$.  Note that with respect to our own D3, the D1 is at a distance $y$ away in the extra-dimensions from us. The validity of our computation of the rate requires $y \gtrsim y_{0} \equiv \pi \sqrt{2 \nu_{1} \alpha'} \sim \pi \sqrt{\alpha'}$. For $y$ larger or much larger than $y_{0}$,  the attractive interaction between our own D3 and the D1 as given in (\ref{c-ampli}) or (\ref{annulus-amplit}) will make the separation smaller\footnote{We here ignore issues caused by the relative motion due to the attractive interaction for simplicity.}. So this can give a small $m_{\rm eff}$ even though either term on the right side of (\ref{detectioncd}) can be on the string scale.  In particular, a small tunable magnetic flux $\hat g$ can be useful to tune the parameter $\nu_{1}$ slightly away from $1/2$ from the second equation in (\ref{nu-parameter}).  This can be used in laboratory to adjust the contribution from the first term on the right side of (\ref{detectioncd}) for a given $y \gtrsim \pi \sqrt{\alpha'}$ to give a small $m_{\rm eff}$.  So indeed we have the possibility to have a small $m_{\rm eff}$.  This in turn implies that a detection of the pair production can be possible even for a small laboratory electric field $E$. 

Note that the dependence of the present rate (\ref{pprate-new}) on the applied laboratory electric and magnetic fields is completely different from the known QED charged spinor rate and  the QED charged scalar rate, respectively, which can be read from 
 \cite{nikishov} as
\be\label{QEDrate}
{\cal W}^{(1)}_{\rm spinor} = \frac{(e E) (e B)}{(2\pi)^{2}} \coth\left(\frac{\pi B}{E}\right) \, e^{- \frac{\pi m^{2}}{e E}},\quad
{\cal W}^{(1)}_{\rm scalar} = \frac{(e E) (e B)}{2 (2\pi)^{2}} {\rm csch} \left(\frac{\pi B}{E}\right) \, e^{- \frac{\pi m^{2}}{e E}}.
\ee
It is also different from the following charged vector rate given in \cite{Kruglov:2001cx} as
\be\label{w-bosonr}
{\cal W}^{(1)}_{\rm vector} = \frac{(e E) (e B)}{ 2 (2 \pi)^{2}} \frac{ 2 \cosh \frac{2 \pi B}{E} + 1}{ \sinh \frac{\pi B}{E}} \, e^{- \frac{\pi m^{2}}{e E}},
\ee
even though we know that our rate (\ref{pprate-new}) is due to the massive charged vector modes.  These sharp differences make the present rate (\ref{pprate-new}) unique, which has its origin from the underlying string theory.  So we can use this unique 
dependence to determine if the rate (\ref{pprate-new}) is the one detected in practice. 

In conclusion, if the underlying theory is indeed relevant to our real world and if our 4-dimensional world can be taken as a D3 brane (or a collection of D3 branes), our present study indicates that there is a potential possibility of detecting the open string pair production using the current available laboratory electric and magnetic fields if a D-string happens to be nearby in the directions transverse to us.  To our D3 brane observer, the D1 at a separation $y$ from the D3 is in general invisible and $y$ is along the extra dimensions. If a detection of the rate (\ref{pprate-new}) and its dependence against the applied tunable electric and magnetic fluxes are confirmed, the first implication shall be the existence of extra dimension(s) and moreover this also gives a possible test  of the underlying string theory.  

While the above discussion appears to open up a possibility of detecting the open string pair production, we would like to address certain issues and cautions related to the actual detection before closing\footnote{We thank very much our anonymous referee for raising certain questions which help to clarify various issues and lead to the following extended discussion.}.   For this, let us mention briefly the issue with 
the usual Schwinger pair production in QED. The critical electric field $E_{c}$ applied for detecting the electron/positron pair can be estimated simply by setting the work done by the electric force over a distance of the Compton length $1/m_{e}$, in natural units, equal to the rest energy of the pair as 
\be 
2 e E \frac{1}{m_{e}} = 2 m_{e} \to E_{c} = \frac{m^{2}_{e}}{e} \sim 10^{18}\, {\rm Volt}/m,
\ee
where $m_{e}$ is the mass of positron and $e$ its charge.  The current laboratory limit of the controllable constant electric field is about $ 10^{10}\, {\rm Volt}/m$, eight orders of magnitude smaller than the above critical field\footnote{\label{fnnew} We searched but didn't find any specific information about the strongest constant electric field which can be realized in a laboratory. We  consulted our experimental colleague Zhengguo Zhao and learned that the current laboratory limit for electric field is on the order of $10^{10}\, {\rm Volt}/m$. The strongest direct-current magnetic field generated is on the order  of  $50$ Tesla, see \cite{smf} for example.  This gives $e E\sim e B \sim 10^{- 8} m^{2}_{e}$ with $m_{e}$ the electron mass.}.
This explains why the Schwinger pair production has not been directly observed so far,  though there are efforts using alternating fields and it is expected to detect this pair production, for example, in the next generation of laser sources.  For spin-1/2 fermions, adding magnetic field does not help much. We can understand this from considering a charged particle with spin-S and mass $m_{S}$ in a constant magnetic field $B$ (with zero electric field), for example see \cite{Ferrara:1993sq}, and the energy spectrum is 
\be \label{energyspectrum}
 E^{2}_{S,\, n} = (2n + 1) e B - g_{S} e B\cdot S + m^{2}_{S},
 \ee 
where $n$ denotes the Landau level and $g_{S}$ the gyromagnetic ratio. Note that this spectrum is the weak-field limit of stringy one (\ref{OSM}) by taking, for example,  $\hat g' = 0$ and $\hat g = 2\pi \alpha' e B \ll 1$. For the lowest Landau level $n = 0$,  we always have $E_{1/2, \,0} = m_{1/2}$  for $S = 1/2$ since $g_{S} = 1/S$ due to the so-called minimal coupling.  
While for a charged massive vector boson, say, W-boson, we have $E^{2}_{1,\, 0} = - e B + m^{2}_{1}$ for which we have $g_{S} = 2$, due to the non-minimal coupling since $g_{S} \neq 1/S$.  If we take large $B/E$ for the rates in (\ref{QEDrate}) and (\ref{w-bosonr}),  to leading order approximation, we find a universal rate formula
\be
{\cal W}^{(1)}_{S} =   \frac{(e E) (e B)}{(2\pi)^{2}} e^{- \frac{\pi E^{2}_{S, 0}}{e E}},
\ee
where $E_{S, 0}$ is given by (\ref{energyspectrum}) with $n = 0$ for $S = 0, 1/2, 1$, respectively, all with $g_{S} =  2$.  Here the requirement of large $B/E$ is consistent with the equation (\ref{energyspectrum}) which holds only in the absence of electric field, i.e. $B/E \to \infty$.  For large $B/E$, adding magnetic field enhances the rate exponentially only for the charged massive vector bosons, say, W-bosons.  In practice, however, this doesn't help much since the laboratory magnetic field, see footnote (\ref{fnnew}), is too small while the W-boson mass $m_{W} \sim 80$ GeV is too large.

The story for the D3/D1 system under consideration is, however,  different.  The D1 acts effectively as a stringy magnetic field to the D3 and gives an intrinsic $\nu_{1} = 1/2$ in the absence of magnetic field on D3.  If we take the D3 as our (1 + 3)-dimensional world, we have in general $\nu_{1} \sim 1/2$ since our laboratory magnetic field cannot be large.  To avoid the tachyon condensation and at the same time to have a possibility of detecting the pair production,   we need to have the brane separation 
$y \gtrsim y_{0}  = \pi \sqrt{\alpha'}$. Once again, note that the ends of the open strings appear to the brane observer as the charged massive particles such as the charged massive vector bosons.  The mass of these vector bosons is $m = y/(2\pi \alpha') \gtrsim M_{s} /2 >$ TeV with $M_{s} = 1/\sqrt{\alpha'}$ the string scale being larger than a few TeV \cite{Berenstein:2014wva}.  However, relevant to the pair production is not this mass but the effective one as given in (\ref{detectioncd}) when we take $\nu_{1} \sim 1/2$. This $m_{\rm eff}$ can in principle be as small as one wishes. This would appear to imply the existence of a window for detection.  Concretely,  for the detection,  we also need to have from (\ref{detectioncd}),
\be\label{detection}
m^{2}_{\rm eff} \sim e E \le 10^{-8} m^{2}_{e},
\ee
which gives the upper bound for $m$ as
\be\label{ub}
m^{2} - \frac{1}{4 \alpha'} \le 10^{-8} m^{2}_{e} \to  m  = \frac{y}{2 \pi \alpha'} \le \frac{M_{s}}{2} \left(1 + 2 \times 10^{-8} \left(\frac{m_{e}}{M_{s}}\right)^{2}\right).
\ee
We then have the brane separation $y$ in the range
\be\label{y-range}
\frac{\pi}{M_{s}} \lesssim y \le \frac{\pi}{M_{s}}  \left(1 + 2 \times 10^{-8} \left(\frac{m_{e}}{M_{s}}\right)^{2}\right).
\ee
Even if we take $M_{s}$ as a few TeV,  we have from the above $y \sim \pi/M_{s} \sim 10^{-19}$ m since $m_{e}/M_{s} \sim 10^{- 7}$.  In other words, once the brane separation is on the order of $10^{-19}$ m, assuming $M_{s}$ on the order of a few TeV, there would appear to
have a possibility detecting the pair production given the current available electric and magnetic fields in laboratory.  The above discussion would appear to give one's impression that the detection of the pair production should not be difficult if all the underlying assumptions are met. In the following, we will explain that this is not so if we put this in actual practice.

To have the actual detection, we first need to have the presence of a  D1 nearby our own D3 along the transverse directions with the brane separation $y  \gtrsim y_{0}  = \pi \sqrt{\alpha'}$.  Note  $y_{0} \le 10^{-19}$ m since $M_{s} \ge$ a few TeV.  Note also that this $y$ is the shortest distance between the D1 and our D3, giving rise to the lightest open strings.  On the one hand, this is good for the experimental setup. For example,  the laboratory sizes for having the electric and magnetic fields as well as for detecting, say, the electric current produced by the pair production are all macroscopic and can therefore all be taken as infinite in comparison with $y \gtrsim y_{0} \le 10^{-19}$ m.  So these setups approximate very well our computations.  However, on the other hand, the probability for having a nearby D1 at a separation $y \gtrsim y_{0} \le 10^{-19}$ m at present time should be very slim. The reason is simple. If string theories are indeed relevant to our real world, we expect that there are plenty of D1 nearby our D3 at the early time of our universe.  In the absence of applied fluxes on our own D3 and the D1, the interaction between the D3 and the D1 can be obtained from (\ref{annulus-amplit}) by turning off the fluxes and setting $\bar \nu_{0} = 0, \nu_{1} = 1/2$ to give
\bea\label{annulus-amplit-noflux}
\Gamma_{3, 1} = \frac{2 V_{2} }{8 \pi^{2} \alpha'} \int_{0}^{\infty} \frac{d t}{t^{2}} ~e^{- \frac{y^{2} t}{2 \pi \alpha' }} \frac{[\cosh \frac{\pi t}{2} - 1]^{2}}{\sinh\frac{\pi t}{2}} \prod_{n = 1}^{\infty} \frac{|1 - 2 |z|^{2n}  \cosh \frac{\pi t}{2}  + |z|^{4n} |^{4}}{(1 - |z|^{2n})^{6} [1 - 2 |z|^{2n} \cosh \pi  t + |z|^{4n}]},
\eea  
which is always positive, therefore giving rise to an attractive interaction (we also reached this same conclusion from the closed string cylinder amplitude given earlier). Note that the above integrand blows up  for $t \to \infty$ when $y < \pi \sqrt{\alpha'}$, indicating a tachyonic instability to occur there.  So the D-strings are attracted towards the D3. Whenever the separation $y$ reaches $y = \pi\sqrt{\alpha'}$, the tachyon condensation occurs and the D-strings dissolve themselves to form the so-called (D3, D1) non-threshold bound state. From the viewpoint of the D3,  the dissolved D1 give rise to now the worldvolume magnetic flux.  

Even if we are lucky enough to have a nearby D1 to our D3,  the actual chance to detect the lightest open string pairs in a laboratory is also very slim.   To explain this in a way easy to follow, let us hide the spatial direction along the D1. So the D1 now looks like a point and our D3 like a plane.  The lightest open string pairs are the ones connecting the point and the plane.  The applied electric field in a laboratory is able to pull apart the virtual charged and  anti-charged pair, being the ends of the virtual open string pair, to make the pair become real such that a detection is possible only when the ends of the lightest virtual open string pair fall in the spatial region of the applied field.  The possibility for this is also extremely slim since the laboratory size, though being macroscopic, is vanishingly small in comparison with the infinitely extended D3.  Furthermore, if we are super-lucky to meet all the above, we also need to be in the right moment when the brane separation $y \sim y_{0}  = \pi \sqrt{\alpha'}$ to be able to detect the pairs produced. Once this moment passes, the tachyon condensation occurs due to $y < y_{0}$, given the attractive interaction between the D1 and the D3.   

In summary, in order to detect the pair production, we need to have the presence of a nearby D1,  our laboratory needs to be in the correct location and the measurement needs to be performed in the right time.  Meeting all these is expected in general to be very difficult and therefore the detection of the pair production should not be an easy task. How to overcome these  experimentally to give a sensible practical detection is beyond the scope of this note.


\section*{Acknowledgments}
The author would like to express his sincere thanks to the anonymous referee for the questions raised which help to clarify certain issues on the actual detection of the pair production. He also thanks his experimental colleague Zhengguo Zhao for information on the laboratory electric field limit and acknowledges the support by a grant from the NSF of China with Grant No: 11235010.

\end{document}